\documentclass{aastex63}

\usepackage{hyperref,epsfig,graphicx,lineno,longtable, amssymb,multirow,amsmath}

\received{***}
\revised{***}
\accepted{***}

\shorttitle{Magnetic Field Strength and CME Dynamics}
\shortauthors{Sun et al.}
\graphicspath{{./}{figures/}}

\begin{document}

\title{How Magnetic Field Strength Affects Stellar Coronal Mass Ejection Dynamics}

\author[0000-0001-5657-7587]{Zheng Sun}
\affiliation{School of Earth and Space Sciences, Peking University, Beijing 100871, People's Republic of China; huitian@pku.edu.cn}
\affiliation{State Key Laboratory of Solar Activity and Space Weather, National Space Science Center, Chinese Academy of Sciences, Beijing 100190, China}
\affiliation{Leibniz Institute for Astrophysics Potsdam, An der Sternwarte 16, Potsdam 14482, Germany}

\author[0000-0001-5052-3473]{Julián D. Alvarado-Gómez}
\affiliation{Leibniz Institute for Astrophysics Potsdam, An der Sternwarte 16, Potsdam 14482, Germany}

\author[0000-0002-1369-1758]{Hui Tian}
\affiliation{School of Earth and Space Sciences, Peking University, Beijing 100871, People's Republic of China; huitian@pku.edu.cn}
\affiliation{State Key Laboratory of Solar Activity and Space Weather, National Space Science Center, Chinese Academy of Sciences, Beijing 100190, China}

\begin{abstract}
Observations show that stellar coronal mass ejection (CME) candidates display relatively lower kinetic energies compared to expectations from solar flare-CME relations extrapolated to the stellar regime. This behaviour was predicted by studies of magnetic confinement of CMEs by strong large-scale stellar magnetic fields. However, the possible promoting role of stronger small-scale magnetic fields has not yet been properly explored in previous studies. In this work, we present the first parametric study that simultaneously incorporates both the promoting and confining effects of magnetic field strength on CME dynamics. We perform CME simulations with scaled solar magnetograms spanning $\langle B_\star\rangle=1,5,10,50,100\ B_\odot$ and inserting flux ropes whose magnetic energy is set to scale as $E_{\rm FR} \propto \langle B_\star\rangle^2$. Our results show that CME speed and mass increase with magnetic field strength in this restrictive scenario, approximately following $v_{\rm CME} \propto \langle B_\star\rangle$ and $M_{\rm CME} \propto \langle B_\star\rangle^{1.5}$. These trends indicate that, within this idealized solar-scaled framework, increasing the magnetic field strength enhances the net promoting forces relative to the confining forces and drives faster, more massive CMEs. We further identify the upward Lorentz force as the dominant contributor to the acceleration and the mass enhancement. We also conducted additional cases with different flux rope energies that do not follow the above scaling assumption, and found that stronger flux ropes produce faster and more massive CMEs for each given stellar model. The adopted scaling assumptions are intended as a controlled parametric experiment rather than a realistic model of young solar-type stars, and future work using more realistic stellar magnetic maps will be required to determine which regions of the parameter space explored here are most relevant to active stars.

\end{abstract}


\section{Introduction} \label{sec:intro}
Coronal mass ejections (CMEs) are large-scale expulsions of magnetized plasma from the Sun into the heliosphere \citep{2011LRSP....8....1C, 2012LRSP....9....3W}, which  could trigger severe space weather effects. CMEs are capable of compressing planetary magnetospheres and driving atmospheric escape \citep{2016IAUS..320..409A, 2020JGRA..12527639G}. When CMEs carry embedded southward magnetic fields, they efficiently couple with Earth's magnetosphere, often initiating geomagnetic storms \citep{2024SoPh..299...93S, 2025GeoRL..5214040F, 2025MNRAS.536.1089H, 2025ApJ...982..194P}. 

CMEs can originate from both active regions (ARs) and quiescent regions, the latter typically involving filament eruptions without accompanying flares \citep{2011LRSP....8....1C}. In this study, we focus specifically on AR-associated CMEs (i.e., flare-related CMEs). The pre-eruption configuration of a CME is typically considered to be a magnetic flux rope (FR), which often forms along polarity inversion lines (PILs) where opposite magnetic polarities converge \citep{2012A&A...539A.131K, 2019LRSP...16....3T, 2020RAA....20..165L, 2023ApJ...953..148S,2024ScChE..67.1592L,2025SCPMA..6879611Z}. Free magnetic energy and helicity for the flux rope eruption accumulate through processes such as flux emergence or shearing motions \citep{2020RAA....20..165L,2024A&A...686A.148S,2025ApJ...990...45S}. Once initiated, the flux rope propagates mostly radially outward, evolving into a CME. The speed of a CME is governed by multiple factors, including gravity, the strength of the overlying strapping field, the amount of stored free energy, the ambient coronal density, and the background solar wind speed \citep{2022RvMPP...6....8S,2022SoPh..297...59K, 2024ApJ...964..159L}. If the upward force on the flux rope is insufficient to overcome gravity and the overlying magnetic field confinement, the eruption would fail to escape, resulting in a confined eruption \citep{2003ApJ...595L.135J,2018Natur.554..211A, 2019ApJ...881..151L}. 

CMEs are not limited to the Sun, but they are also expected to occur on other cool stars \citep{2007AsBio...7..167K, 2007AsBio...7..185L}. In recent years, the search for stellar CMEs has been a popular area of research \citep{2024ApJ...966...24Y, 2024Univ...10..313V,2025ApJ...985..219X}, with some candidate events reported. For example, \citet{2021NatAs...5..697V} used observations from XMM-Newton, Chandra, and EUVE to identify dimming events on cool stars associated with stellar flares. These dimmings are interpreted as stellar analogs of solar coronal dimming, i.e., regions of reduced emission caused by plasma evacuation after CME eruptions \citep{2025arXiv250519228V}.
Observations at optical wavelengths have also been used in detecting potential filament eruptions, particularly through the identification of red-blue asymmetries in Doppler-shifted spectral lines (e.g., \citealt{1990A&A...238..249H, 1997A&A...321..803G, 2022NatAs...6..241N, 2022A&A...663A.140L, 2025ApJ...978L..32L, 2023ApJ...948....9I}). Similar asymmetries have been reported in ultraviolet \citep{2011A&A...536A..62L,2026NatAs..10...64N} and X-ray spectral lines \citep{2019NatAs...3..742A,2022ApJ...933...92C,2024ApJ...969L..12I}. Another promising method involves detecting transient absorption features in the background X-ray emission, potentially caused by CMEs crossing the line of sight to the emitting region (e.g., \citealt{1999A&A...350..900F, 2017ApJ...850..191M}). Moreover, stellar analogs of Type II and Type IV solar radio bursts, which could be associated with CMEs, have also been observed \citep{2020ApJ...905...23Z,2025Natur.647..603C,2025A&A...703A.198K}.

Previous statistical studies of solar flares and CMEs have demonstrated that more energetic flares are generally associated with faster and more massive CMEs \citep{yashiro2008statistical, 2009IAUS..257..233Y,2019SSRv..215...39L}. The most powerful solar flare on record, the Carrington event in 1859, had an estimated bolometric energy of approximately $10^{33}$ erg \citep{2021ARA&A..59..445H}. In contrast, some active stars are capable of producing superflares with energies reaching up to $10^{37}$ erg \citep{2024LRSP...21....1K}. Motivated by this large energy gap, \citet{2013ApJ...764..170D} first extended the solar flare–CME relationship to active stars. They found that the associated CME energy and mass loss would be unrealistically large for such energetic and frequent stellar flares, suggesting that the flare–CME relationship must flatten for active stars. \citet{2017MNRAS.472..876O} further applied the solar flare–CME scaling laws to estimate mass loss from active stars, and the predicted CME mass loss rates were significantly overestimated compared to observational constraints. This discrepancy suggests that further explanations are needed to account for why CME mass loss from active stars is lower than expected. \citet{2018ApJ...862...93A} first used numerical simulations to show that strong stellar magnetic fields can effectively confine CMEs. They found that the CME speeds were significantly reduced, leading to a corresponding decrease in kinetic energy. In line with these simulations, \citet{2019ApJ...877..105M} used existing observational data to show that many stellar CME candidates have kinetic energies approximately two orders of magnitude lower than expected based on solar scaling. Further modeling by \citet{2022MNRAS.509.5075S} of torus instability in force-free magnetic fields revealed that the onset of instability is significantly more difficult on stars with strong large-scale magnetic fields, thereby reducing the likelihood of CME eruptions.

While previous studies have emphasized the confining effects of strong magnetic fields on CMEs, they have often overlooked their promoting roles. Active regions with stronger magnetic fields could provide more magnetic energy for the eruption \citep{2003GeoRL..30.2181V}. Moreover, stars with stronger magnetic fields are typically associated with faster background stellar winds \citep{2009ApJ...699..441V,2023MNRAS.524.5060C}, which can further enhance CME acceleration. In addition, magnetic reconnection in the stronger magnetic fields may provide more free energy for the CME \citep{2000mare.book.....P,2017SoPh..292...17D,2025arXiv250817208W}.
Thus, when evaluating the influence of strong stellar magnetic fields on CMEs, it is essential to consider both their confining and promoting effects—a duality that has not been thoroughly addressed so far. In this work, we aim to isolate how magnetic field strength influences CME dynamics under controlled solar-scaled assumptions. Rather than attempting to reproduce realistic young solar-type stars, we construct an idealized parameter space in which the promoting and confining effects of magnetic fields can be systematically compared. In the following, we present our methodology and simulation setup in Section~\ref{sec:methods}, the results in Section~\ref{sec:results}, and the discussion and conclusions in Section~\ref{sec:discussion}.

\section{Methods} \label{sec:methods}
\subsection{Magnetic Maps} \label{sec:maps}
\begin{figure}
    \centering
    \plotone{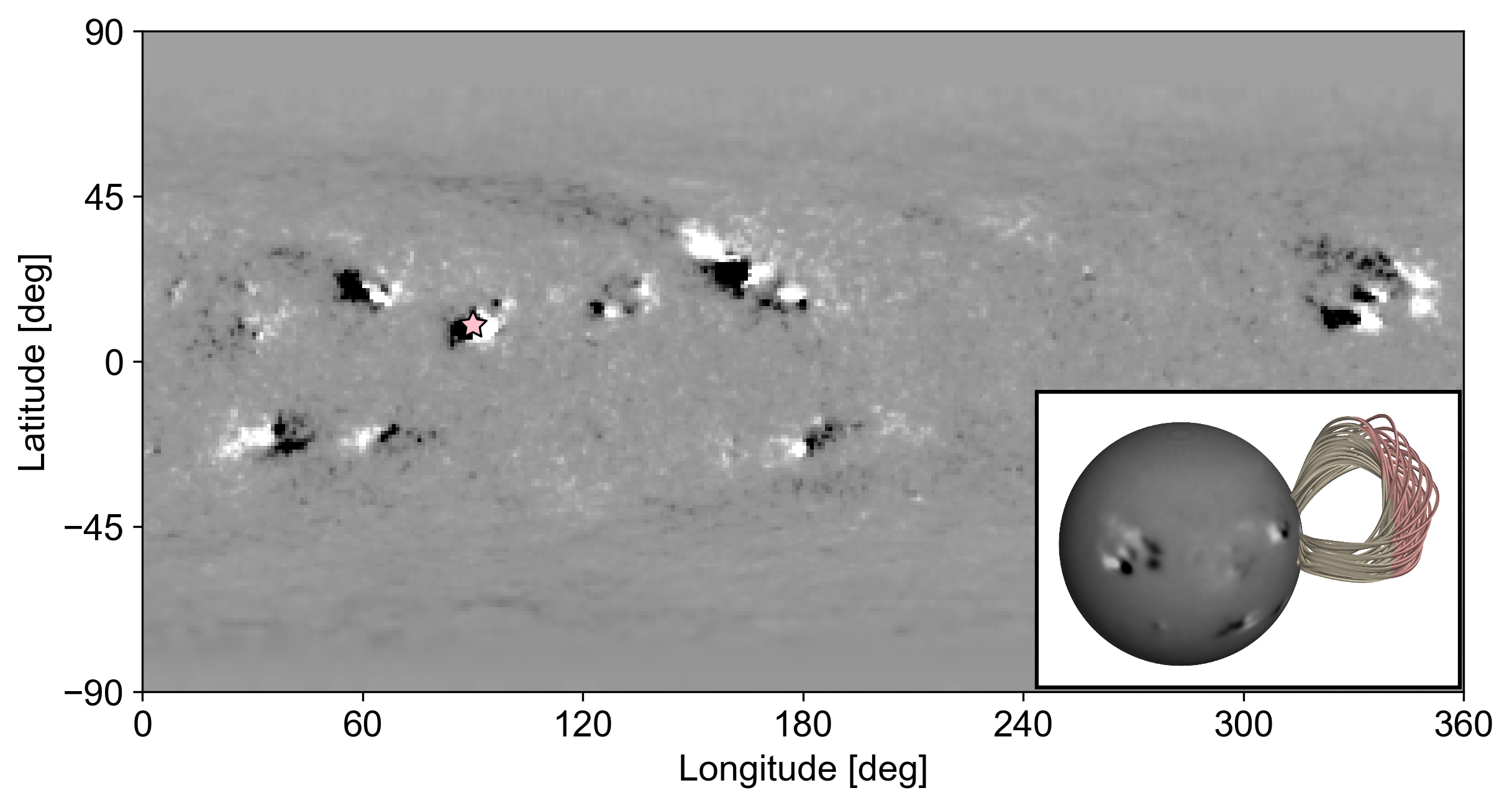}
  \caption{Magnetic field maps for the five stellar models, scaled by factors of 1, 5, 10, 50, and 100, corresponding to average surface magnetic field strengths of $\langle B_\star \rangle = 1, 5, 10, 50, 100\ B_\odot$, respectively. The pink star marks the location where the flux rope is inserted. The lower-right inset shows a 3D visualization of the magnetic map and the inserted flux rope. The pink segment of the flux rope indicates the region used later to calculate the forces acting on the flux rope.
    \label{fig:Fig.1}}
  \end{figure}

First, we establish a set of scaled solar models that span a range of magnetic field strengths, from weak to strong.
To achieve this, we adopt a single solar magnetogram and scale it to produce the required global photospheric magnetic field maps, a method that has also been employed in previous studies (e.g., \citealt{2020IAUS..354..426J,2022ApJ...938....7L}).
Specifically, we use the synoptic magnetogram from Carrington Rotation (CR) 2107, corresponding to a period near solar maximum, as shown in Figure \ref{fig:Fig.1}. The average radial magnetic field $B_r$ in this map is approximately 2.8 G. We define this value as one solar magnetic unit, denoted as $B_\odot$. To construct stellar models with enhanced magnetic fields, we scale this map by factors of 5, 10, 50, and 100, resulting in stellar models with photospheric average $B_r$ values of 1, 5, 10, 50, and 100 $B_\odot$, respectively. The purpose of these scaled maps is not to reproduce realistic stellar magnetic topologies, but to isolate the effect of magnetic field strength within an idealized framework in which the AR geometry and magnetic topology are kept unchanged across all models.

\subsection{MHD Models}

We perform our simulations using the Space Weather Modeling Framework (SWMF; \citealt{2005JGRA..11012226T}), which was originally developed for modeling various components of the solar system, including the solar corona, planetary magnetospheres, and the heliosphere \citep{2018LRSP...15....4G,2021JSWSC..11...42G}. In recent years, SWMF has been successfully extended to stellar environments beyond the solar system (e.g., \citealt{2011ApJ...733...67C,2011ApJ...738..166C,2014MNRAS.438.1162V,2015MNRAS.449.4117V,2016A&A...588A..28A,2016A&A...594A..95A,2020ApJ...902L...9A,2016ApJ...833L...4G, 2022ApJ...941L...8G}). In this work, we primarily utilize the Alfvén Wave Solar Model (AWSoM; \citealt{2014ApJ...782...81V}) module, which serves as the Solar Corona (SC) component of SWMF. AWSoM solves the three-dimensional magnetohydrodynamic (MHD) equations using the Block-Adaptive-Tree Solar-wind Roe-type Upwind Scheme (BATS-R-US; \citealt{1999JCoPh.154..284P}) in spherical coordinates. This model incorporates the propagation, reflection, and dissipation of Alfvén waves, which contribute to both coronal heating and stellar wind acceleration.

Our stellar CME simulation consists of two main steps. First, using the magnetic maps as input of AWSoM, we construct steady-state stellar wind models to simulate the background stellar corona. Using the five scaled magnetic map in Section~\ref{sec:maps}, we construct five steady-state stellar models with $\langle B_\star\rangle=1,5,10,50,100\ B_\odot$.
Throughout this study, we constrain our models to the regime of solar-type stars, assuming the stellar mass and radius are identical to those of the Sun. To isolate the effect from magnetic field strength, we also fix the rotation period and source surface height to solar values—25.4 days and 2.5 R$_{\star}$, respectively. Since observations indicate that stars with stronger magnetic fields often rotate faster \citep{2009ARA&A..47..333D,2012LRSP....9....1R,2017AN....338..428K}, and the source surface height for other stars might be different \citep{2022MNRAS.514.5465W}, we revisit these two parameters in Section \ref{sec:discussion}.
The Alfvén wave correlation-length parameter and the Alfvén wave Poynting-flux-to-magnetic-field ratio are adopted as commonly used solar values in AWSoM, namely $L_\perp\sqrt{B}=1.5\times10^5\ {\rm m\,T^{1/2}}$ and $S_A/B=1.1\times10^6\ {\rm W\,m^{-2}\,T^{-1}}$, respectively. The simulation domain extends from 1 $R_\star$ to 60 $R_\star$, with angular resolutions of 1.4° in both latitude and longitude, and a minimum radial grid size of approximately 0.0007~$R_\star$. The grid resolution is fixed for both the steady-state and time-dependent CME simulations. At this resolution, large-scale features of the solar magnetic field, such as active regions, and large coronal structures, are well-resolved and can be accurately represented. 
Smaller-scale features may be under-resolved, but we do not expect them to affect the CME propagation much.
In the second step, we insert a flux rope into the steady-state models to initiate the CME. The parameters of the flux ropes will be introduced in Section \ref{sec:fluxrope}.

\subsection{Flux Ropes} \label{sec:fluxrope}

The energy used to build up flux ropes are thought to be from primarily two mechanisms: flux emergence and sheared photospheric motions \citep{2018SSRv..214...46G}. However, observational evidence for flux rope emergence remains limited, with only a few candidates reported (e.g., \citealt{2005ApJ...622.1275L, 2008ApJ...673L.215O}). Therefore, here we focus on flux rope formation via sheared motions.

In this scenario, shearing motions at the photosphere converting the potential magnetic loops into sheared loops. These sheared loops may subsequently reconnect to form a coherent flux rope structure and erupt \citep{2001ApJ...552..833M,2021NatAs...5.1126J}. The magnetic field lines of the resulting flux rope originate from the loops in the AR, implying that the magnetic field strength of the flux rope should be proportional to that of the AR:

\begin{equation}
B_{\rm FR} \propto B_{\rm AR}.
\end{equation}

Assuming that the geometric properties of flux ropes are similar, and noting that magnetic energy scales as $E =\int B^2/8\pi\ dV$, the magnetic energy of the flux rope $E_{\rm FR}$ should scale as:

\begin{equation}
E_{\rm FR} \propto B_{\rm FR}^2.
\end{equation}

From Section \ref{sec:maps}, we assume that the AR field strength scales with the stellar average magnetic field $\langle B_\star \rangle$, leading to the key scaling relation:

\begin{equation}
E_{\rm FR} \propto B_{AR}^2 \propto \langle B_\star\rangle^2.
\end{equation}

Notably, although $B_{\rm AR} \propto \langle B_\star\rangle$ holds in our idealized models, this relation is not yet strongly supported by observations. Therefore, this assumption should be understood as part of our idealized scaling framework rather than as an observationally established property of active stars. The possible discrepancy between this assumption and realistic stellar magnetic properties is further discussed in Section~\ref{sec:discussion}. Based on these assumptions, we construct a set of CME simulations on the stars with $\langle B_\star \rangle$=1, 5, 10, 50, and 100 $B_\odot$. We use Gibson-Low (GL) ﬂux rope model \citep{1998ApJ...493..460G} to initiate the CME, which has been used in many previous studies (e.g., \citealt{2004JGRA..109.1102M,2004JGRA..109.2107M,2017ApJ...834..173J,2020IAUS..354..426J,2018ApJ...862...93A}).
For the solar case, we insert a flux rope with an energy of $E^0_{\rm FR } = 2\times10^{31}$ erg, representative of a relatively weak solar CME. For other models with $\langle B_\star \rangle$=5, 10, 50, and 100 $B_\odot$, we insert flux ropes with energies scaled accordingly:
$E_{\rm FR} = (5^2, 10^2, 50^2, 100^2) \times E^0_{\rm FR}$.
Flux ropes are inserted at the location marked by the green star (91° longitude, 10°N latitude) in Figure \ref{fig:Fig.1}, with an orientation of 7° (matching the underlying main PIL for this AR). 

These simulations allow us to incorporate both the confining and promoting effects of magnetic fields. The promoting effect associated with increased magnetic energy in the flux rope is applied manually, based on the equation (3). All other promoting and confining effects are self-consistently included in the 3D MHD time-dependent evolution captured by our models.
Above all, our idealized stellar models rely on two main assumptions: (i) Because we employ scaled solar magnetograms as input, we assume that the magnetic configurations and properties remain identical across the stars (e.g., the ratio $B_{\rm AR}/\langle B_\star\rangle$). (ii) We assume that the magnetic energy of stellar flux ropes scales with the square of the AR magnetic field strength, implying that the the formation mechanism of flux ropes is similar. Later in Section \ref{sec:discussion} we will discuss the validity of these assumptions for solar-type stars.

\section{Results} \label{sec:results}

\begin{figure}
    \centering
    \plotone{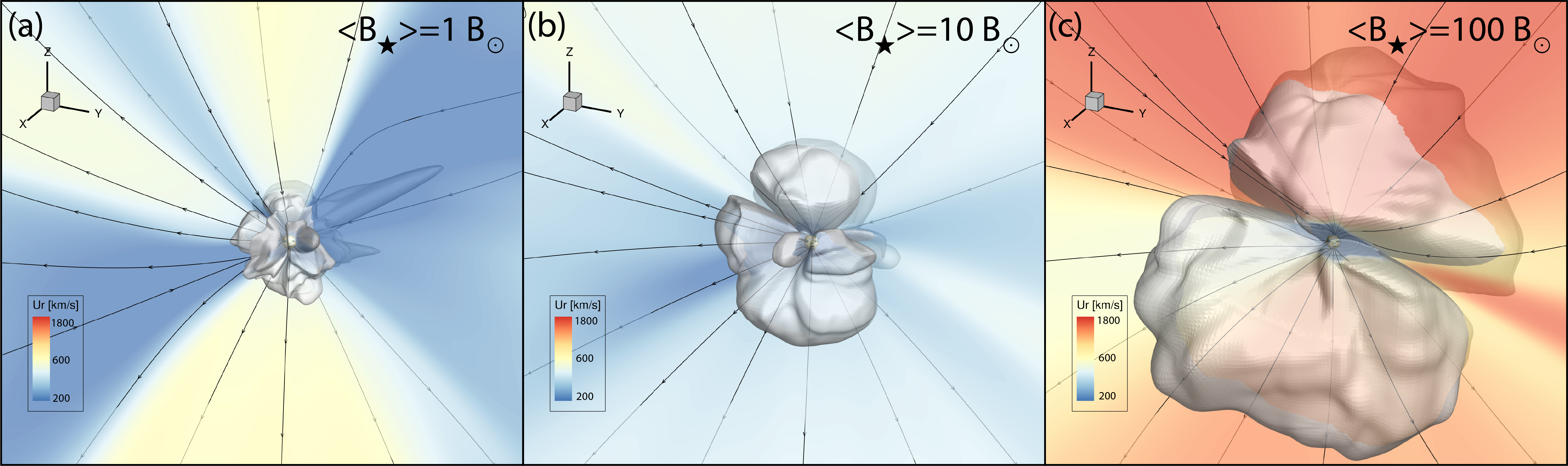}
  \caption{Snapshots of the steady-state stellar wind solutions for models with average surface magnetic field strengths of $\langle B_\star \rangle = 1\ B_\odot$ (a), $10\ B_\odot$ (b), and $100\ B_\odot$ (c). The gray isosurface represents the Alfvén surface.
    \label{fig:Fig.2}}
  \end{figure}

Figure \ref{fig:Fig.2} presents the steady-state stellar wind models for the $\langle B_\star \rangle = 1$, 10, and 100~$B_\odot$ cases (models with $\langle B_\star \rangle = 5$ and 50~$B_\odot$ are not shown in the figure). The results demonstrate that stronger stellar magnetic fields lead to faster stellar winds. For example, in the $\langle B \rangle = 1\ B_\odot$ case, the wind speed ranges from 100 to 800 km$\cdot$s$^{-1}$, whereas in the 100 $B_\odot$ case, it ranges from 500 to 1800 km$\cdot$s$^{-1}$. This indicates that stars with stronger magnetic field can provide greater acceleration from stellar wind, which belongs to the promoting effect of strong magnetic field.
In addition, the size of the Alfvén surface increases with magnetic field strength, with average radii of 8, 14, and 32~$R_\star$ for the $\langle B_{\star} \rangle =$1, 10, and 100~$B_\odot$ models, respectively. These trends are consistent with previous simulating studies (e.g., \citealt{2014MNRAS.438.1162V,2023MNRAS.524.5060C,2023MNRAS.524.2042E}). We also calculated the mass loss rates of the steady-state models. The case with $\langle B_\star \rangle = 1\ B_\odot$ yields a mass loss rate of $2.5\times10^{-14}\ M_\odot\mathrm{yr}^{-1}$, which is consistent with the typical range observed for the solar mass loss rate \citep{2019ARA&A..57..157C}. The other two cases give mass loss rates of $8.2\times10^{-14}\ M_\odot\mathrm{yr}^{-1}$ and $3.3\times10^{-13}\ M_\odot\mathrm{yr}^{-1}$, respectively. Compared with the mass-loss estimates inferred from Ly$\alpha$ astrospheric absorption for solar-type stars with similar magnetic field strength \citep{2002ApJ...574..412W,2003ApJ...598.1387P}, these values are lower by approximately one order of magnitude. This discrepancy may result from the fact that our model adopts the solar rotation period, rather than the shorter rotation periods typical of young solar-type stars. Another possible explanation is the choice of Alfvén wave parameters in AWSoM (here we just use the solar values), which are known to have a strong impact on the mass-loss rate predicted by the model \citep{2020A&A...635A.178B}.

\begin{figure}
    \centering
    \plotone{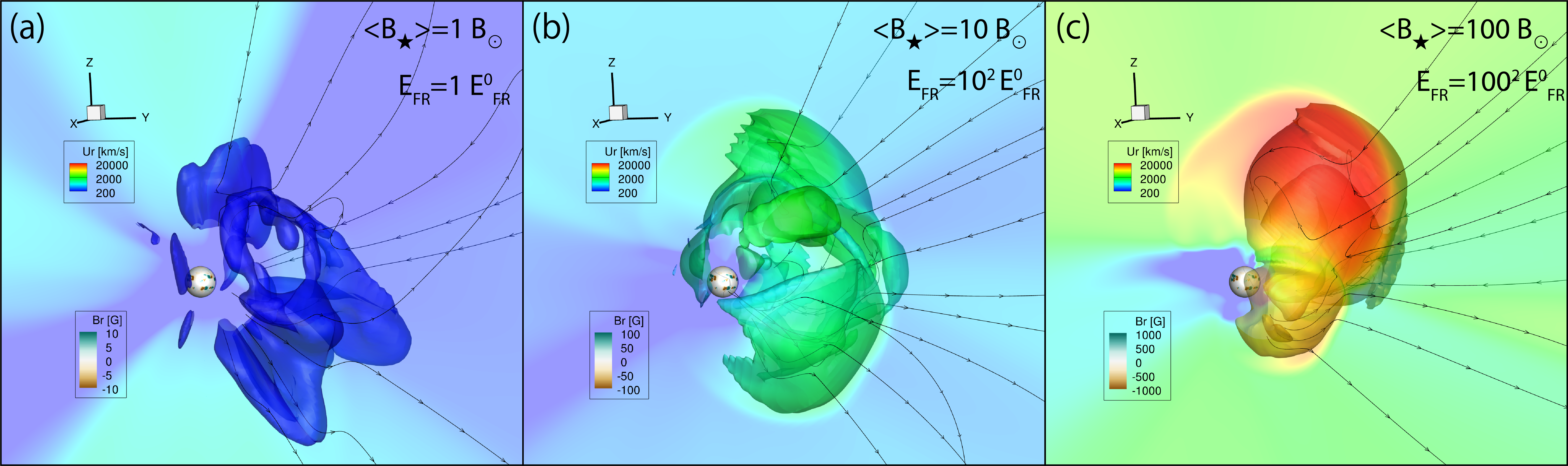}
  \caption{Snapshots of CMEs in models with average surface magnetic field strengths of $\langle B_\star \rangle = 1\ B_\odot$ (a), $10\ B_\odot$ (b), and $100\ B_\odot$ (c). The CME front is defined as the region where the density exceeds three times the steady-state background value, i.e., $n/n_0 > 3$. Each panel corresponds to the moment when the CME front has propagated to a radial distance of $10\ R_\star$, occurring at approximately 180, 36, and 4.5 minutes after the eruption onset for panels (a), (b), and (c), respectively. (An accompanying animation shows the time evolution of the CME propagation for the three models from eruption onset to this stage. Because the propagation times differ substantially among the models, each panel in the animation uses its own timestep sequence.})
    \label{fig:Fig.3}
  \end{figure}

Figure \ref{fig:Fig.3} shows snapshots of CMEs in the $\langle B_{\star} \rangle = 1$, 10, and 100~$B_\odot$ models. 
To quantify CME speeds, we follow the method of \citet{2020ApJ...895...47A} and first identify the CME bulk as those surface points where the density exceeds three times the steady‐state background value ($n/n_0>3$). We then define the leading edge by selecting the top 10\% of these points in terms of radial distance. The propagation speed is measured at a heliocentric distance of $R = 10\ R_\star$, as the majority of reported solar CME speeds are derived from LASCO observations within \(\lesssim30\\ R_\odot\) \citep{2009EM&P..104..295G,2024arXiv240704165G}. We then follow the criterion in \citet{2024ApJ...971..153X} to calculate the CME mass at $R = 10\ R_\star$.

\begin{figure}
    \centering
    \plotone{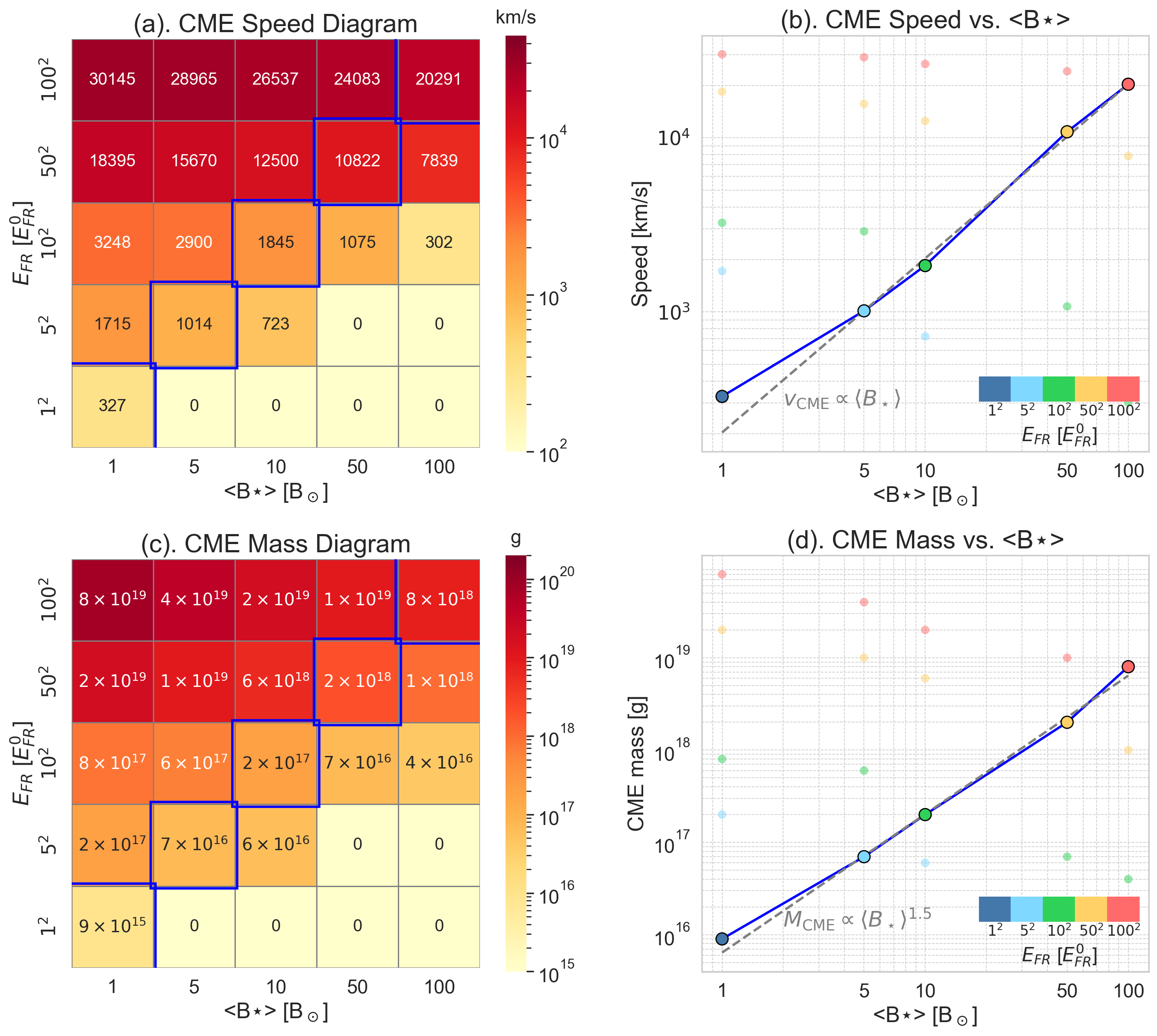}
\caption{Scaling of CME speed and mass with stellar magnetic field strength and flux rope energy.  (a) Heat map of CME speed as a function of $B_\star$ and $E_{\rm FR}$; blue-framed points along the diagonal correspond to models with $E_{\rm FR} \propto \langle B_\star \rangle^2$.  (b) CME speed versus $B_\star$ extracted from the diagonal of (a); the gray dashed line is a linear fit, showing $v_{\rm CME} \propto \langle B_\star \rangle$.  (c) Heat map of CME mass as a function of $B_\star$ and $E_{\rm FR}$, with the same diagonal models highlighted.  (d) CME mass versus $B_\star$ extracted from the diagonal of (c); the gray dashed line shows $M_{\rm CME} \propto \langle B_\star \rangle^{1.5}$. Off-diagonal cases are shown as smaller scattered points.}  \label{fig:Fig.4}
    \end{figure}

Figure \ref{fig:Fig.4}(a) shows a heat map of the calculated CME speeds in each scaled-magnetogram model with different flux rope energies. The diagonal (outlined in blue) marks the cases satisfying equation (3), where the flux rope energy scales with $\langle B_{\star} \rangle^2$. The first, third, and fifth pixels on this diagonal correspond to results of Figures \ref{fig:Fig.3}(a), (b), and (c), respectively. Figure \ref{fig:Fig.4}(b) displays the CME speeds as a function of $\langle B_{\star} \rangle$, showing a approximate linear relationship $v_{\rm CME} \propto \langle B_\star \rangle$ for our diagonal events following our assumed scaling relation. We note that the most extreme diagonal case reaches a CME speed of approximately 20,000~km$\cdot$s$^{-1}$, which is likely unrealistically large. This result arises directly from our idealized scaling assumptions, in which the inserted flux ropes preserve similar geometrical properties and levels of non-potentiality while their magnetic free energy increases proportionally to $\langle B_\star\rangle^2$. We also tested stricter CME-identification thresholds of $n/n_0>5$ and $n/n_0>8$ for the diagonal cases, and found that the resulting CME speed scaling remains approximately linear with $\langle B_\star\rangle$, indicating that the main trends are not sensitive to the specific threshold choice.
We also perform additional simulations by varying the magnetic energy of the flux ropes, deviating from the values expected in equation (3). The corresponding results are shown in the elements above and below the diagonal.
Cases with a speed of zero indicate that the CME front does not reach $R = 10 R_\star$. Since the Alfvén surface of most our models lies beyond this distance—and in the solar case it is about $8\ R_\star$, very close to $10\ R_\star$—we interpret CMEs that fail to propagate to $10\ R_\star$ as being confined events. 
A horizontal examination of the heatmap indicates that, for fixed flux rope energy, stronger background fields produce slower CMEs, consistent with the magnetic suppression scenario \citep{2018ApJ...862...93A}. A vertical view of the heatmap shows that within a given background field, higher flux rope energy leads to faster CMEs. For the first column (the solar regime), the data follow a power-law relation of $v_{CME} \propto 10^{2.53} E_{FR}^{0.50}$. Compared with \citet{2017ApJ...834..173J}, we find that the CME speeds in our simulations increase with the flux rope energy more rapidly than in their study. However, when focusing only on the $\langle B_{\star} \rangle =$1 $B_\odot$ and $\langle B_{\star} \rangle =$5 $B_\odot$ cases (which are closer to solar conditions), the increase rate is comparable to theirs. Later in this section, we will discuss why the rate in these two cases differs from that in the other cases.
Panels (c) and (d) show the similar analysis of the CME masses. For the diagonal cases satisfying equation (3), the CME mass follows a quasi–power-law scaling, $M_{\rm CME} \propto \langle B_\star \rangle^{1.5}$. For other cases, they show a similar trend with the CME speed:
Horizontally, stronger background magnetic fields with the same flux rope energy correspond to less massive CMEs. However, this trend is not consistent with the results of \citet{2018ApJ...862...93A}, who found that the CME mass remains nearly constant. Vertically, within a given background field, cases with higher flux rope energy produce more massive CMEs, and for the solar regime, the relation is approximately described by $M_{CME} \propto 10^{15.90} E_{FR}^{0.99}$. 

To understand the physical mechanisms driving the variation in CME speeds and masses across different models, we analyze the specific forces acting on the CME. The promoting effects associated with stronger magnetic fields include: (i) an AR with a stronger magnetic field and a flux rope with higher magnetic energy (ii) greater acceleration from faster stellar winds, and (iii) enhanced magnetic reconnection.
The first effect is manifested in the increased Lorentz force acting on the flux rope. The acceleration due to stellar wind is reflected in the dynamic pressure of the background wind. Magnetic reconnection influences CME propagation through two main mechanisms: dynamic pressure from reconnection outflows, and enhancement of the toroidal magnetic flux within the CME \citep{2018SoPh..293..113W}. The latter is incorporated into the Lorentz force acting on the CME.
The confining effect primarily arises from the magnetic tension force exerted by the overlying field. Additionally, thermal pressure must be considered, though its dependence on magnetic field strength remains uncertain. By combining all these contributions, we evaluate the net force acting on the flux rope to assess the interplay between promoting and confining effects:

\begin{equation}
\mathbf{f}_{\rm net} = \mathbf{f}_{\rm Lorentz, FR} + \mathbf{f}_{\rm strapping} + \mathbf{f}_{\rm wind} + \mathbf{f}_{\rm thermal} + \mathbf{f}_{\rm reconnection}+ \mathbf{f}_{\rm g}.
\end{equation}

Here, $\mathbf{f}_{\rm Lorentz, FR}$ is the Lorentz force acting on the flux rope, given by $\mathbf{f}_{\rm Lorentz, FR}=\mathbf{J}_{FR} \times \mathbf{B}_{FR}$. $\mathbf{f}_{\rm strapping}$ is the magnetic tension force from the overlying magnetic field, generally directed downward, and can be represented by $\mathbf{f}_{\rm strapping}=(\mathbf{B}_{0} \cdot \nabla)\mathbf{B}_0/4\pi$, where $\mathbf{B}_0$ is the background steady-state magnetic field. $\mathbf{f}_{\rm wind}$ represents the dynamic pressure force from the stellar wind, expressed as $\mathbf{f}_{\rm wind}=\rho (\mathbf{v}_{\rm wind} - \mathbf{v}_{\rm CME})^2$. When the wind speed exceeds the CME speed, this term acts as a promoting force; otherwise, it acts as a drag force. $\mathbf{f}_{\rm thermal}$ denotes the thermal pressure gradient force, $\mathbf{f}_{\rm thermal}=-\nabla p$. $\mathbf{f}_{\rm reconnection}$ represents the dynamic pressure force from reconnection outflows, given by $\mathbf{f}_{\rm reconnection}=\rho_{\rm outflow} \mathbf{v}_{\rm outflow}^2$ \citep{2010RvMP...82..603Y}. $\mathbf{f}_{\rm g}$ is the gravitational force, which has been shown to be negligible when the magnetic field strength exceeds 30 G \citep{2004SoPh..219..169L,2005IAUS..226..250R}. Given that the magnetic fields considered in our models are significantly stronger than this threshold, the relative contribution of gravitational force becomes even smaller; therefore, we neglect $\mathbf{f}{\rm _g}$ in our analysis (although it is captured in the behaviour of our time-dependent models). 

\begin{figure}
    \centering
    \plotone{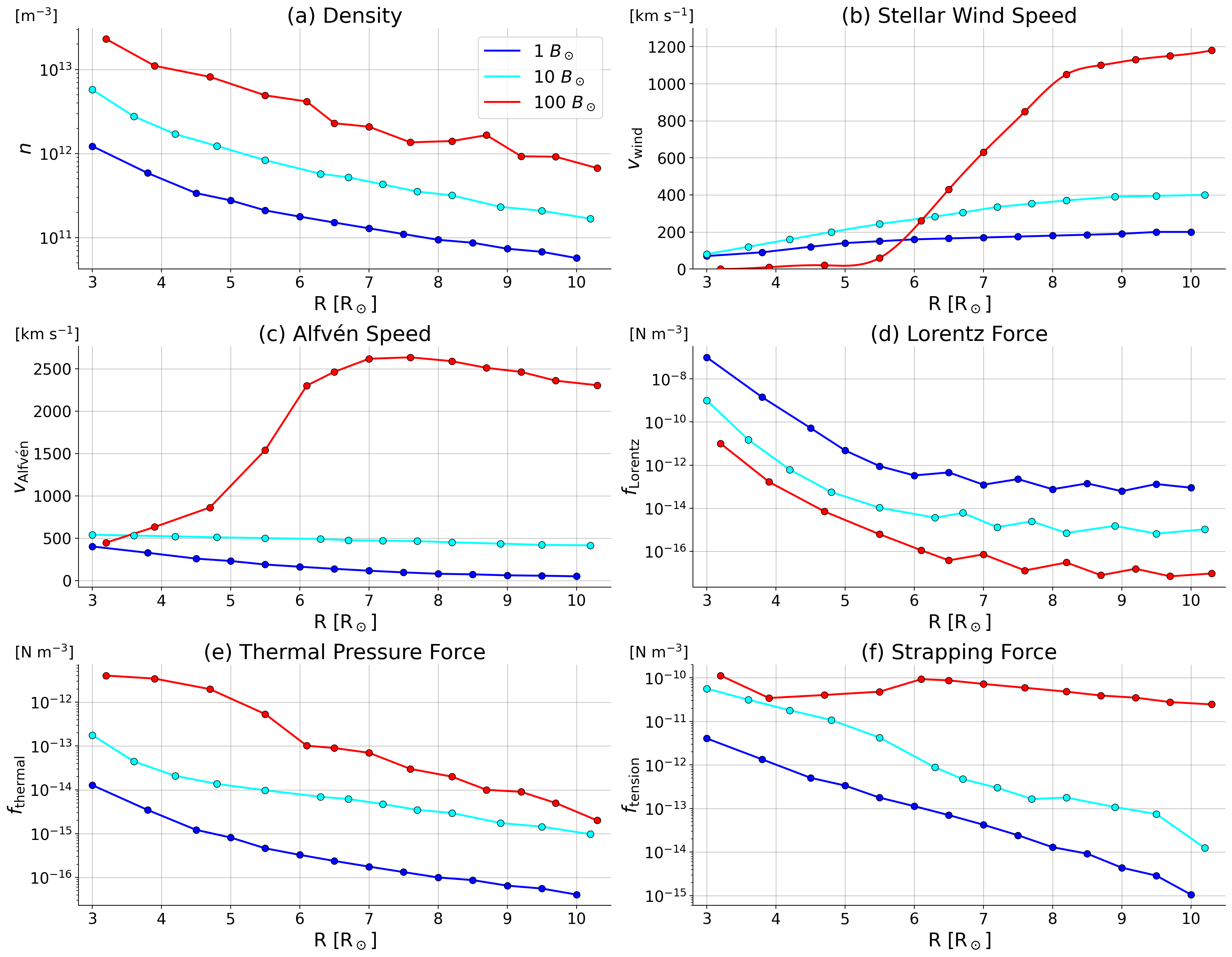}
  \caption{Parameter distributions along the CME propagation for models with $\langle B_\star \rangle = 1$, $10$, and $100\ B_\odot$. The profiles are taken along the radial line corresponding to the direction indicated by the green star symbol in Figure 1. Panels show the following quantities: (a). CME front density, (b) stellar wind speed, (c) Alfvén speed, (d) Lorentz force, (e) thermal pressure force, (f) Strapping force.    \label{fig:Fig.5}}
  \end{figure}

We select three representative models with $\langle B_\star\rangle = 1, 10, 100$ G for a detailed force analysis. First, we examine the variation of the average density as the CME front propagates to $10~R_\star$, as shown in Figure \ref{fig:Fig.5}(a). It is evident that cases with stronger magnetic fields exhibit higher densities; however, when the magnetic field strength increases by one order of magnitude, the corresponding density increase is less than an order of magnitude. Figures \ref{fig:Fig.5}(b–f) then show the temporal evolution of various forces (or related parameters) from the onset of the CME until the front reaches $10R_\star$. The CME front is defined as the outermost 10\% of points (in terms of radial distance) among those with density enhancements of $n_{\rm ratio}>3$. We verified that, in all cases explored here, the identified fronts generally coincide well with the main propagating CME bulk and with the regions of maximum outward propagation speed. Similar density-threshold-based approaches have also been widely adopted in previous CME simulation studies to identify CME ejecta and derive propagation properties (e.g. \citealt{2017ApJ...834..173J,2018ApJ...862...93A,2020ApJ...895...47A}). We track the positions of these points in the steady-state wind solution to obtain an average stellar wind speed (panel (b)), thermal pressure force (panel (e)), and strapping force (panel (f)) acting on the CME. For the Lorentz-force analysis (panel (d)), instead of using the density-defined CME front, we directly track the magnetic flux rope itself. Immediately after the flux-rope insertion, we trace the magnetic field lines rooted at the flux-rope footpoints and select the outermost 25\% portion of these field lines in radial distance (Illustrated in Fig. \ref{fig:Fig.1}). At later times, we continue tracing the same magnetic footpoints and again select the outermost 25\% portion of the corresponding field lines. The Lorentz force density is then calculated over this evolving magnetic-flux-rope volume. To evaluate the influence of reconnection outflows—whose speed is expected to be comparable to the local Alfvén speed in the current sheet—we calculate the Alfvén speed in the current sheet beneath the flux rope. Specifically, we define a conical region with its apex at one-third of the CME height and its base at the CME front, and compute the average Alfvén speed within this cone, as shown in Figure \ref{fig:Fig.5}c. We consider this range approximately corresponds to the height of the current sheet \citep{2019MNRAS.482..588Y}.
The kinetic energy of the CME is given by $ E_k = \tfrac{1}{2}\rho v^2 = \int f \, dr$. In our calculations, the path length of the integration is consistently set to $10~R_\star$ across all models. The CME density, $\rho$, increases with magnetic field strength, while the CME speed approximately follows $v_{CME} \propto B_{\star}$ in the cases meeting the equation (2). This implies that when the magnetic field strength increases by one order of magnitude, there must exist a dominant force term that increases by more than two orders of magnitude. In the following, we analyze which force plays this dominant role.

We first examine $\mathbf{f}_{\rm wind}$. Panel (b) of Figure \ref{fig:Fig.5} shows the radial stellar wind speed at the location of the CME front. In the 100~$B\odot$ case, the wind speed reaches $\sim$1200 km s$^{-1}$ at $R = 10 R_\star$, which is significantly lower than the CME speed of $\sim$20,000 km s$^{-1}$. In the 10~$B_\odot$ model, the stellar wind speed is $\sim$400 km s$^{-1}$, also much smaller than the CME speed of $\sim$1800 km s$^{-1}$. This indicates that once the CME speed exceeds the stellar wind speed, the wind can no longer provide additional acceleration and instead will exert drag on the expanding structure. Therefore, $\mathbf{f}_{\rm wind}$ cannot account for the observed CME acceleration. It should be noted that, in the solar case, the wind speed reaches only about 200 km s$^{-1}$ at 10 $R_\odot$, which is lower than the typical solar wind speed at this height. This is because the region corresponds to the streamer belt, where the plasma outflow is intrinsically slow due to the closed magnetic topology and high plasma density.
We next consider $\mathbf{f}_{\rm reconnection}$. According to fast reconnection theory, the reconnection outflow speed should be comparable to the local Alfvén speed \citep{2000mare.book.....P,2010RvMP...82..603Y}. Panel (c) shows that in the 100~$B\odot$ model the Alfvén speed reaches $\sim$2500 km s$^{-1}$, while in the 10~$B_\odot$ model it is $\sim$500 km s$^{-1}$—both far below the CME speeds. Thus, the contribution from $\mathbf{f}_{\rm reconnection}$ is also negligible.
Next, we examine $\mathbf{f}_{\rm strapping}$. As shown in panel (d), the strapping force density increases with $B$, scaling approximately as $\mathbf{f}_{\rm strapping} \propto \langle B_\star\rangle^2$. However, since this force acts opposite to the CME propagation direction, it serves to confine rather than accelerate the CME.
We then analyze $\mathbf{f}_{\rm thermal}$. Figure \ref{fig:Fig.5}(e) shows that the thermal pressure force density varies by less than two orders of magnitude among the 1, 10, and 100~$B\odot$ models. Consequently, $\mathbf{f}_{\rm thermal}$ cannot account for the large differences observed in CME speeds.
Finally, we consider $\mathbf{f}_{\rm Lorentz,FR}$. Panel (f) shows the Lorentz force density at the CME front. We find that when $B$ increases by one order of magnitude, $\mathbf{f}_{\rm Lorentz,FR}$ increases by more than two orders of magnitude, which can explain the observed difference in CME velocities. Since no other force term exhibits a comparable scaling, we conclude that $\mathbf{f}_{\rm Lorentz,FR}$ is the dominant contributor to CME acceleration in this study. It should be noted that the purpose of this analysis is primarily to compare how each force term scales among different stellar magnetic field strengths. Therefore, the relative variations of the same force term across different models are more meaningful than the absolute comparison between different force terms, since the corresponding quantities are evaluated over different physical regions.

In cases with $\langle B_{\star} \rangle \geq 10B_\odot$, the CME speed exceeds both the stellar wind speed and the local Alfvén speed. Therefore, the contributions from wind pressure and reconnection outflows can be neglected. In contrast, in the solar model (the $\langle B_{\star} \rangle =$1 $B_\odot$ model), where $v_{\rm CME}$ is comparable to both the stellar wind and Alfvén speeds, additional acceleration from wind and reconnection outflows produces a slight deviation above the linear scaling trend (seen in Figure \ref{fig:Fig.4} (a) and (c)). This explains why the curve between first two points appear flatter, consistent with \citet{2017ApJ...834..173J}.

\section{Discussion} \label{sec:discussion}

In this work, we use scaled solar magnetograms with $\langle B_\star\rangle = 1, 5, 10, 50, 100\ B_\odot$ and insert flux ropes whose magnetic energy is set to scale as $E_{\rm FR} \propto \langle B_\star\rangle^2$ to conduct a series of CME simulations. We find that CME speed and mass increase approximately as $v_{\rm CME} \propto \langle B_\star\rangle$ and $M_{\rm CME} \propto \langle B_\star\rangle^{1.5}$ when the energy of the flux rope is scaled as above in this idealized regime. There is also observational support from solar studies. Statistical analyses have shown that CME speed correlates with the magnetic flux of the source ARs \citep{2003GeoRL..30.2181V, 2018SoPh..293...60M}, consistent with our findings. Empirical solar studies also suggest $M_{\rm CME}\propto E_{\rm flare}^{0.6}$ \citep{2012ApJ...760....9A,2013ApJ...764..170D}. Assuming $E_{\rm flare}\propto B^2$, this implies $M_{\rm CME}\propto B^{1.2}$, broadly consistent with our simulated scaling.

The simulations presented here represent a systematic parameter study designed to isolate the role of magnetic field strength rather than to reproduce specific solar or stellar events. By scaling the solar magnetograms and flux rope energies, we explore how the magnetic field modifies the competition between the forces that promote eruption and those that confine it. This approach provides physical insight into the underlying processes that regulate CME dynamics in magnetically active environments. Despite its advantages, this framework involves several idealized simplifications that may introduce deviations from realistic stellar magnetic conditions on young solar-type stars. Specifically, it assumes (i) magnetic configurations identical to those of the Sun, and (ii) a scaling of flux rope energy with the AR magnetic field strength as $E_{\rm FR} \propto \langle B_{\rm AR}\rangle^2$. These assumptions define the limits of applicability of our model, and their validity is examined in the following discussion.

The first simplification is that the stars possess magnetic configurations identical to those of the Sun. In our models, the AR magnetic field strength scales proportionally with the global field strength, resulting in a constant ratio of $B_{\rm AR}/B_\star$. However, some current observations suggest that this ratio actually decreases as the global field strength increases. For the Sun, AR magnetic fields are typically at the kilo-Gauss level, whereas the large-scale field is only a few Gauss, implying $B_{\rm AR}/B_\star \sim 10^2$--$10^3$ \citep{1994ApJ...425L.117P,2012ApJ...757...96D,2020Sci...369..694Y,2024Sci...386...76Y,2026ApJ..1000..205Q}. In contrast, observations of active stars may suggest kilo-Gauss magnetic fields in both their large-scale background fields and starspot regions \citep{2018A&A...616A.160V,2019A&A...626A..86S,2020ApJ...893..113P,2024A&A...691L..12M}, implying that $B_{\rm AR}/B_\star$ may decrease to values of only a few to $\sim1$--$10$. This suggests that ARs on young solar-type stars may not be able to form flux ropes with energies as large as those adopted in our models. In this case, the corresponding stellar CMEs would likely occupy the off-diagonal regions below the diagonal sequence in Figures~\ref{fig:Fig.4}(a) and \ref{fig:Fig.4}(c), where the confining effect becomes increasingly important while the promoting effect no longer increases proportionally. As a result, the corresponding CME speeds and masses would be substantially lower than those predicted by the idealized diagonal scaling relations. This trend is consistent with the magnetic suppression scenario \citep{2018ApJ...862...93A, 2019ApJ...884L..13A}. In addition, active stars may possess magnetic topologies substantially different from the solar case. Zeeman-Doppler Imaging (ZDI) observations suggest that young active stars can exhibit strong dipolar or quadrupolar large-scale fields, different multipolar complexity, high-latitude or polar active regions, and larger spot coverage than the Sun \citep{1989A&A...225..456S,1997A&A...326.1135D,2001ApJ...551.1099S,2017AN....338..753F}. These topology differences may significantly modify the balance between promoting and confining effects by changing the overlying strapping field, the distribution of open magnetic flux, and the locations and sizes of eruptive active regions. Moreover, active stars may exhibit larger ARs than the Sun, and their latitudinal distribution of ARs could differ \citep{2001ApJ...551.1099S,2017AN....338..753F}. Some observations suggest that young solar-type stars may host strong ARs at high latitudes, including the polar regions \citep{1989A&A...225..456S,1997A&A...326.1135D}, though a recent study indicates that the distribution of flaring ARs may still resemble that of the Sun \citep{2025A&A...695A..21Y}. While these factors could influence CME dynamics, they are beyond the scope of this study and will be addressed in future work.

The second simplification concerns the scaling of the flux rope energy, $E_{\rm FR} \propto \langle B_\star\rangle^2$. This assumption is based on the idea that flux ropes have similar twist numbers across different stars, which reflects the comparable levels of non-potentiality in their ARs. The main challenge, however, is whether the non-potential parameters—such as the $B^2$-normalized helicity or the ratio of free to total magnetic energy—are indeed similar to those of the Sun. For instance, if shearing motions on active stars are slower than solar values, the resulting helicity injection rate would be reduced, making it difficult to build flux ropes with energies as high as those assumed in our models \citep{2024ESS.....520303J}. This could lead to large deviations from the scaling relations presented in this study.
A further concern is the possible existence of an upper limit on magnetic helicity. Theoretical studies suggest that force-free magnetic fields may reach a helicity upper limit $H_{\rm ul}$, beyond which plasma ejection occurs in the form of a CME to reduce the helicity \citep{2006ApJ...644..575Z,2008ApJ...683.1160Z}. If active stars possess a lower ratio of helicity to magnetic energy, i.e., $H_{\rm ul}/B^2$, than in the solar case, the formation of flux ropes with the assumed energy scaling may be inhibited \citep{2015ApJ...804L..28S,2016ApJ...828...62J,2017A&A...601A.125P}. Addressing this issue would require further theoretical analysis or observational constraints from solar ARs.
Moreover, our assumption is based on the scenario in which flux ropes are formed through sheared photospheric motions, as discussed in Section \ref{sec:fluxrope}. However, simulation studies have demonstrated that flux ropes can also form through flux emergence (e.g., \citealt{2008A&A...492L..35A,2009ApJ...697.1529F}). In such cases, the magnetic energy content of the emerged flux rope is more difficult to constrain, as it may depend on the underlying stellar dynamo and convective processes. A detailed investigation of flux-emergence-driven flux ropes is therefore beyond the scope of this study.

The parameters adopted in the AWSoM model, such as the Alfvén wave Poynting-flux-to-magnetic-field ratio and the Alfvén wave correlation length, remain uncertain for stars other than the Sun. Previous studies have attempted to constrain them observationally through stellar mass-loss rates and X-ray observations \citep{2022ApJ...924..115O,2025ApJ...995...83C}, as well as HST measurements \citep{2020A&A...635A.178B,2021ApJ...916...96A}, while some other stellar applications of AWSoM have adopted solar values for these parameters (e.g., \citealt{2022ApJ...941L...8G,2022ApJ...928..147A,2023MNRAS.524.5060C}). In the present work, we also adopt the solar values in order to isolate the role of magnetic field strength and explore the relative behavior of CME dynamics within a controlled parameter space. Among the various model parameters, we here focus on two quantities that are more directly connected to the large-scale magnetic configuration and may potentially influence the CME dynamics: the stellar rotation period and the source surface height.
Stronger magnetic fields are generally associated with shorter stellar rotation periods \citep{2014MNRAS.441.2361V,2018MNRAS.474.4956F,2022A&A...662A..41R}. To evaluate the effect of rotation on our idealized models, we re-ran the $\langle B_\star\rangle = 1$, 10, and $100\,B_\odot$ cases using a rotation period of 1 day instead of the solar value of 25.4 days. We found that the resulting CME properties remain very similar, with differences within approximately 5\%, indicating that stellar rotation does not significantly affect the main conclusions of this study within the explored parameter space.
The source surface height, $R_{\rm s}$, defines the layer above which magnetic field lines are assumed to be open, while the region below may exert strong confinement \citep{2022MNRAS.514.5465W}. In stellar applications, $R_{\rm s}$ is often taken to be $2$--$6\ R_\star$ \citep{2014MNRAS.438.1162V,2015MNRAS.449.4117V,2017MNRAS.466.1542S}. To further evaluate its influence, we performed additional simulations with different source surface radii for the diagonal $\langle B_\star\rangle=100\,B_\odot$ case. We find that although increasing $R_{\rm s}$ enhances the confining effect from the overlying magnetic field, the resulting CME properties remain only weakly affected because the promoting Lorentz-force contribution still dominates within our idealized scaling framework. The detailed results are presented in Appendix~\ref{sec:appendix_rs}. These tests suggest that variations in the source surface height do not significantly alter the main conclusions of this parameter study.

Current observational constraints on stellar AR magnetic fields and magnetic non-potentiality remain highly uncertain. As a result, the physically relevant parameter space for stellar CMEs is still poorly constrained. The present work therefore adopts a broad range of magnetic field strengths and flux rope energies in order to explore how CME dynamics respond to different magnetic assumptions within a controlled framework.
Observations show that most reported stellar CME candidates exhibit projected speeds of only a few hundred km s$^{-1}$ (e.g., \citealt{2019A&A...623A..49V,2022NatAs...6..241N}), while only a small number of events reach speeds above $10^3$ km s$^{-1}$ (e.g., \citealt{1990A&A...238..249H,2023ApJ...948....9I,2025Natur.647..603C}). By contrast, the most extreme case in our idealized diagonal sequence reaches a speed of $\sim10^4$ km s$^{-1}$. In addition, observational estimates of stellar mass-loss rates suggest that the CME masses and velocities of active stars may be lower than those implied by our diagonal cases \citep{2013ApJ...764..170D,2017MNRAS.472..876O,2002ApJ...574..412W,2021ApJ...915...37W,2026arXiv260400925W}. Taken together, current observations may be more consistent with regions of the parameter space in which the promoting effect does not increase as rapidly as assumed along the diagonal sequence, corresponding to models below the diagonal in Figure~\ref{fig:Fig.4}. 
Nevertheless, the full stellar CME parameter space remains largely unconstrained. The model grid presented here therefore provides a useful framework for interpreting future observations and for identifying which combinations of magnetic field strength and flux rope energy are capable of producing different eruptive regimes. Also, if highly energetic CMEs do occur on active stars, their large dynamic pressures may strongly compress planetary magnetospheres and enhance atmospheric erosion, potentially reducing the long-term habitability of exoplanets around magnetically active stars.

Future extensions of this framework should employ more realistic stellar magnetic configurations derived from observations and dynamo models. In particular, ZDI magnetograms, combined with constraints from rotationally modulated optical, ultraviolet, and X-ray observations, may provide more realistic large-scale topologies and AR distributions for young solar-type stars. In addition, stellar dynamo simulations may offer more physically motivated estimates of the magnetic energy and non-potentiality of stellar flux ropes. Such developments will help determine how the trends identified in the present idealized parameter study extend to realistic stellar environments.

\section{Conclusion} \label{sec:conclusion}

In this work we construct a comprehensive grid of CME models to incorporate both the promoting and the confining effects of magnetic field strength on CME dynamics within a single framework. Using scaled solar magnetograms with $\langle B_\star\rangle=1,5,10,50,100\ B_\odot$ and inserting flux ropes whose magnetic energy is set to scale as $E_{\rm FR} \propto \langle B_{AR}\rangle^2$, we conducted a series of CME simulations. Within the adopted scaling assumptions, we find that CME speed and mass increase with magnetic field strength, approximately follows $v_{\rm CME} \propto \langle B_\star\rangle$ and $M_{\rm CME} \propto \langle B_\star\rangle^{1.5}$. These trends indicate that in our idealized models, increasing the magnetic field strength enhances the net promoting terms relative to the confining terms, resulting in greater acceleration. We further find that the upward Lorentz force is the primary contributor to this acceleration. We also conducted additional cases with different flux rope energies that do not follow the above scaling assumption. For a given stellar model, we found that a larger flux rope energy leads to a faster and more massive CME.

The models constructed here adopt two idealized simplifications: (i) the stars possess solar-like magnetic configurations, and (ii) the flux rope energy scales with the AR magnetic field strength as $E_{\rm FR} \propto \langle B_{AR}\rangle^2$. 
These simplifications enable us to isolate the influence of magnetic field strength on CME dynamics but should be kept in mind when applying the results to young solar-type stars. In particular, (i) the ratio $B_{\rm AR}/B_\star$ may vary among different stars, and future observations of stellar magnetic fields are needed to better constrain this relationship; and (ii) non-potential parameters, such as the $B^2$-normalized helicity and the ratio of free to total magnetic energy, may differ from solar values, potentially altering the CME properties. Current observations provide only limited constraints on stellar AR magnetic fields, magnetic non-potentiality, and stellar CME properties. As a result, the physically relevant stellar CME parameter space remains poorly constrained. The present framework is therefore intended as a broad parameter-space exploration that isolates the physical consequences of different magnetic-field assumptions, rather than an attempt to reproduce a specific stellar population. Future work will build on this framework to better understand CME dynamics in young solar-type stars. 

\acknowledgments
This work is supported by the National Natural Science Foundation of China (12425301 \& 425B2024), and the Specialized Research Fund for State Key Laboratory of Solar Activity and Space Weather. H.T. is also supported by the New Cornerstone Science Foundation through the Xplorer Prize. This work was carried out using the SWMF and BATS-R-US tools developed at the University of Michigan’s Center for Space Environment Modeling. We also thank the technical support of ``National Large Scientific and Technological Infrastructure Earth System Numerical Simulation Facility"\footnote{https://cstr.cn/31134.02.EL}.

\bibliographystyle{aasjournal}
\bibliography{ref}

\appendix
\section{Influence of the Source Surface Radius} \label{sec:appendix_rs}

In the main text, we adopt a source surface radius of $R_{\rm s}=2.5\,R_\star$, corresponding to the typical solar value. However, previous studies have suggested that the source surface height of active stars may differ substantially from the solar case and may depend on the stellar magnetic topology, wind properties, and rotation rate \citep{2022MNRAS.514.5465W}. To evaluate the sensitivity of our results to the adopted source surface radius, we performed additional simulations for the diagonal $\langle B_\star\rangle=100\,B_\odot$ case using $R_{\rm s}=2.5$, $7.5$, and $15\,R_\star$.

Figures~\ref{fig:FigA1} (a) and (b) show the steady-state stellar wind density and speed profiles for the three source surface radii. Increasing $R_{\rm s}$ leads to only slightly higher plasma densities and slightly slower stellar winds in the low corona due to the larger fraction of closed magnetic flux and the delayed opening of magnetic field lines to larger heights.

The resulting CME propagation speeds are shown in Figure~\ref{fig:FigA1} (c). Although increasing $R_{\rm s}$ enhances the confining effect from the overlying magnetic field and modifies the stellar wind environment, the resulting CME speeds remain relatively similar among the three cases. Within our idealized scaling framework, the increase in the promoting Lorentz-force contribution is substantially larger than the additional confinement introduced by larger source surface radii. As a result, the main conclusion of this work, namely that the promoting Lorentz force dominates over the confining effects in the strong-field regime, remains unchanged for the explored range of $R_{\rm s}$.
\newpage

\renewcommand{\thefigure}{A.\arabic{figure}}
\setcounter{figure}{0}

 \begin{figure*}
\centering
\plotone{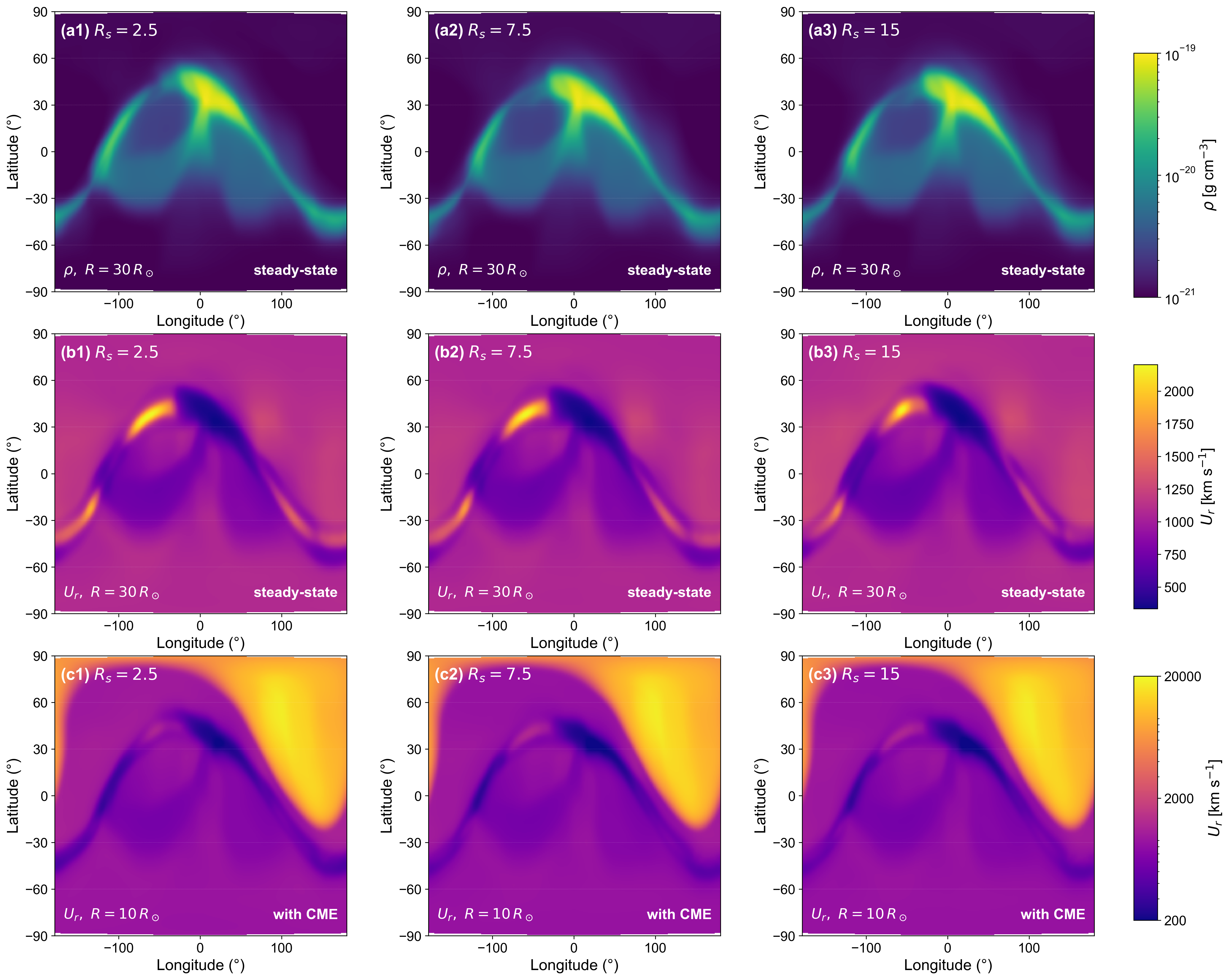}
\caption{Comparison of the $\langle B_\star\rangle =100\ B_\odot$ diagonal case obtained with source surface radii of $R_{\rm s}=2.5$, $7.5$, and $15,R_\star$ (left to right columns). The top row shows the steady-state stellar wind density distributions on the spherical surface at $R=30\ R_\star$, while the middle row shows the corresponding steady-state radial wind velocity distributions at $R=30\ R_\star$. The bottom row shows the radial velocity distributions on the spherical surface at $R=10,R_\star$ when the CME bulk reaches $R=10\ R_\star$. 
\label{fig:FigA1}
}
\end{figure*}

\end{document}